\documentclass[11pt,twoside]{article}
\usepackage{newpasp,epsf,psfig}
\def\edcomment#1{\iffalse\marginpar{\raggedright\sl#1\/}\else\relax\fi}
\reversemarginpar \def\gsim{\lower.5ex\hbox{$\; \buildrel > \over \sim \;$}}

\begin{document}

\title{Revealing the Energetics and Structure of AGN Jets}

\author{Eric S. Perlman\altaffilmark{1}, Herman L. Marshall\altaffilmark{2} \& 
John A. Biretta\altaffilmark{3}}

\altaffiltext{1}{Joint Center for Astrophysics, University of 
Baltimore County, 1000 Hilltop Circle, Baltimore, MD  21250  USA}
\altaffiltext{2}{Center for Space Research, Massachusetts Institute of
Technology, 77 Massachusetts Ave., Cambridge, MA  02139 USA}
\altaffiltext{3}{Space Telescope Science Institute, 3700 San Martin Drive, 
Baltimore, MD  21218 USA}


\begin{abstract}

Until very recently, few constraints existed on the physics of jets, even
though they represent the first known evidence of mass outflow in AGN.  This
has begun to change with HST and Chandra observations, which allow us to
observe short-lived, dynamic features, and compare their spectra and morphology
to those of longer-lived particles seen in the radio.  We examine HST and
Chandra observations of M87 and 3C273 which reveal that these two prototype
objects seem radically different.

\end{abstract}

\section{Introduction}

Collimated jets occur in many settings, including galactic nuclei, binary star
systems, and star formation.  AGN jets are relativistic flows composed of
high-energy particles and magnetic fields,  which emit synchrotron radiation in
the optical and radio.  Jets were the first AGN outflows observed, discovered
by Curtis (1918), who noticed a `curious, straight ray' emanating from M87. 

Prior to the launch of HST, optical emissions had been observed from two AGN
jets: M87 and 3C273 (Schmidt 1963).  ROSAT and {\it Einstein} observed X-ray
emissions from the jets and/or lobes of these (Schreier et al. 1982, Stern \&
Harris 1986)  plus four others: Cen A  (Burns et al. 1983), Cygnus A  (Harris
et al. 1994), 3C390.3 (Harris et al. 1998) and 3C120 (Harris et al. 1999).  
HST observations,  particularly the 3CR Snapshot Survey (Martel et al. 1999),
have drastically increased the number of known X-ray or optical jets to about
20, representing a fair cross-section of radio-loud AGN properties.

Because of the featureless nature of synchrotron spectra, the jump from
morphology to physics is large for jets.   Some progress has been made through
numerical modeling and multi-frequency radio mapping, but this elucidates only
a small part of the energy spectrum, and details can be obscured by the long
particle lifetimes ($10^{5-6}$ yr).   Due to their short radiative lifetimes,
optical and X-ray synchrotron emitters  ($\sim 1-100$ yr), represent much more
dynamic characteristics. Thus to obtain the tightest constraints, multiband
data are required.

\section{An issue of stratification}

One of the surprises provided by HST observations was that the optical and
radio morphologies of jets can be quite different.  For example, in 3C273 the
optical jet appears much narrower than seen in the radio, with a twisted, even
`braided' morphology (R\"oser et al. 1997).  And in M87, there is a large-scale
correspondence of features, but detailed comparison reveals a narrower,
knottier jet in the optical (Sparks et al. 1996).  Therefore it should not have
been a surprise when the first Chandra data  revealed yet more differences. 
For example, in Cen A,  the X-ray and radio maxima of several knots appear
offset by up to an arcsecond (Kraft et al. 2000).  What do these  represent: 
different physical conditions, different emission mechanisms, or both?

Perlman et al. (1999) proposed that the radio-optical differences in the M87
jet could be explained by stratification. What led to this conclusion was HST
and VLA polarimetric images, which revealed large differences in knots, where
the magnetic field vectors seen in the optical (but not radio) become
perpendicular to the jet upstream of flux maxima, followed by sharp decreases
in optical polarization at flux maxima (their Figures 3-6).  Under the Perlman
et al. model, high energy electrons are concentrated along the axis and in
knots, where the magnetic field is compressed by shocks, while the sheath is
dominated by lower-energy particles, with a more static, parallel magnetic
field.  

Naively, the idea of a stratified jet is not novel:  in a fire hose, the
fastest part of the flow is in the center.  In fact, a decade ago some authors
proposed 'two-fluid' jet models  (cf. Sol, these proceedings).  Stratification
has far-reaching implications.   For example, instabilities need not involve
all components of the jet flow.  Moreover, a compressed,  perpendicular
magnetic field in knot regions is a  recipe for particle acceleration.  It
instructive to look at the X-ray and optical flux and spectral morphology,
where several of these features should show up.

\section {Spectral and X-ray Morphology of the M87 and 3C273 Jets}

M87 (d=16 Mpc) and 3C273 (z=0.158) are the prototype jets, having both the
largest angular extents and high surface brightness.  They  are very different
objects, differing by a factor 100 in luminosity and jet power, and by a factor
30 in physical jet length.  One might therefore expect a detailed comparison to
show interesting differences. 

\begin{figure} \plotfiddle{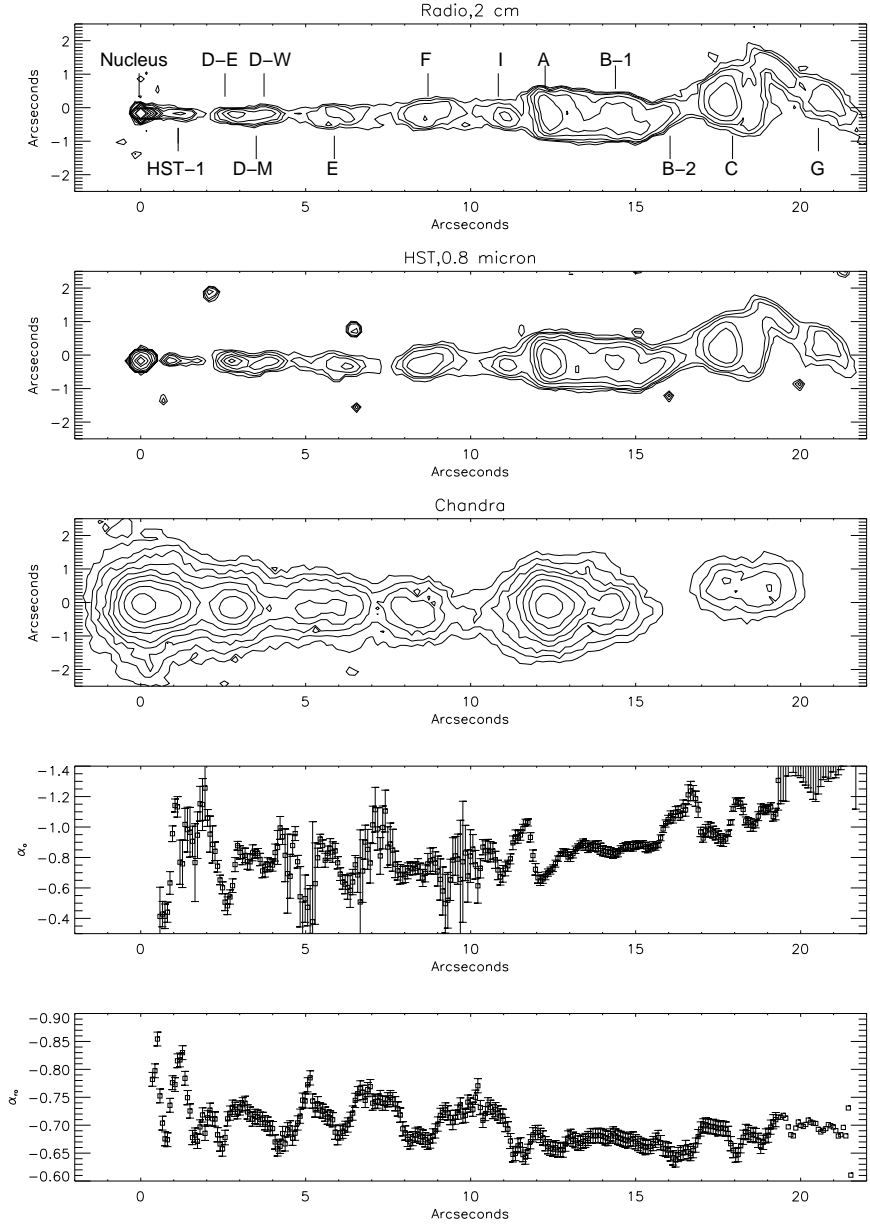}{6.3in}{0}{65}{65}{-200}{-30}

\caption[]{Radio (VLA, 2cm, top), optical (HST, I band, second panel),  X-ray
(Chandra, third panel), $\alpha_o$ (fourth panel)  and  $\alpha_{ro}$ (bottom)
images of  the M87 jet.  The Chandra ACIS+HETG image was adaptively smoothed. 
Contours represent (3, 5, 8, 16, 32, 64, 128, 256...) $\times$ rms noise.  Note
the strong correlation between optical flux and spectral index.  This, along
with polarimetric imaging, strongly argues for energetic stratification.   The
X-ray and optical peaks of most knots are located in the same places, but there
are two optically faint regions where X-ray peaks are seen, in the D-E and E-F
transitions.}

\end{figure}

Looking first at Figure 1, we can see that in the M87 jet, there is an
excellent overall correlation between optical flux and optical spectral index
$\alpha_o$ ($F_\nu \propto \nu^{\alpha}$): in high surface brightness regions, 
one sees flatter spectra.  Near knot maxima one sees exactly the kind of
hardening, followed by softening, that one would expect if particle
acceleration is occurring in the knots.  Interestingly, the optical and
radio-optical $(\alpha_{ro})$ spectral indices do not vary together.  In each
knot region in the inner jet,  we observe  $\alpha_o$ either leading or lagging
$\alpha_{ro}$.  One can understand this in the context of particle acceleration
(Kirk et al. 1999):  if the acceleration timescale is much less than the
cooling timescale for optical synchrotron emitters, one expects $\alpha_o$ to
lead $\alpha_{ro}$; however if the acceleration and cooling timescales are of
similar order, one would expect $\alpha_{ro}$ to lead.  A different situation
is seen in the outer jet, where there is an anticorrelation between $\alpha_o$
and $\alpha_{ro}$.  This might indicate that the jet core has a steeper
injection index than the sheath.

X-ray emission is seen from all regions of the M87 jet, and several optical
and  X-ray knot maxima are located at the same $\theta_{nuc}$, e.g., HST-1
(nuclear distance $\theta_{nuc}=1''$), A ($\theta_{nuc}=12''$) and C
($\theta_{nuc}=18''$).  Knot A's X-ray and optical maxima are at the same
$\theta_{nuc}$, not displaced by $0.5''$, as had been claimed by  Neumann et
al. (1997) and B\"ohringer et al. (2000)  The most likely explanation is the
unexpected brightness of HST-1, which Chandra only partially resolves from the
core, and is hopelessly blended with it in ROSAT and XMM images.  There are,
however,  significant morphological differences.  In knot E
($\theta_{nuc}=5-6''$), the X-ray bright region begins $1''$ upstream of the
optical peak, in an optically faint region.  Also, in knot F
($\theta_{nuc}=8''$), the X-ray bright region begins $0.6''$ upstream of the
optically bright region, and the  maxima are displaced.  The difference in knot
D ($\theta_{nuc}=2.5''-4''$) is more subtle: the decline in X-ray flux
following maximum is much steeper than in the optical.  

The X-ray spectral indices and broadband SEDs indicate that the X-ray emissions
of the M87 jet are due to synchrotron radiation (B\"ohringer et al. 2001,
Marshall et al. 2001a).  Since the lifetimes of X-ray synchrotron emitting
electrons are only a few years, particle acceleration is required.  Spectral
fits and variability timescales constrain these regions to be a small fraction
of the volume of each knot (Harris et al. 1997, Perlman et al. 2001, Marshall
et al. 2001).

Turning to Figure 2, we see that in the 3C273 jet there is no  correlation
between the optical flux and spectral index.  Instead, there is a gradual
steepening in $\alpha_o$ with $\theta_{nuc}$, overlaid with small variations,
e.g.,  flattenings in the A-B1, B1-B2 and C1-C2 transitions.  Thus physical
conditions in the 3C273 jet change remarkably smoothly over scales of many
kiloparsecs.   A detailed spectral index map also reveals evidence of
superposed emission regions in some knots (Jester et al. 2001a).  This is very
different from the M87 jet, illustrated above.

Jester et al. (2001a) note that the consistency of the ground and HST-observed
runs of $\alpha_o$, the gradualness of the spectral changes and the lack of
correlation between optical flux and $\alpha_o$,  are consistent with no
radiative cooling over scales of many kiloparsecs.   But optical synchrotron
emitting particles have lifetimes $\sim$ hundreds of years.  Jester et al.
(2001a) conclude that the only way to escape this paradox is to have continuous
reacceleration, throughout the jet's length. This is not inconsistent with
stratification, but it is very different from what we see in M87.  Optical
polarimetry would be invaluable to examine these issues further. Unfortunately
the current HST polarimetry (Thomson et al. 1993) is too low signal to noise;
reobservation is required.

Comparing optical and X-ray images of 3C273 (Marshall et al. 2001b, Sambruna et
al. 2001), one sees further differences. Knot A is by far the most powerful in
the X-rays, whereas the optical fluxes of all knots are within a factor of two,
and the radio maximum occurs in knot H.  In addition, the X-ray and optical
peaks in knot B appear offset, and no X-ray emission is seen  in knot H2. 
There is debate over the X-ray emission mechanism for knot A:  Marshall et al.
(2001b) and R\"oser et al. (2000) favor synchrotron radiation, while Sambruna
et al. (2001) claim that Comptonization is required.  But these authors agree
that the X-ray emissions of the other knots is due to Comptonization of CMB
photons. Deeper observations are needed to distinguish between mechanisms.

{\centerline{\psfig{file=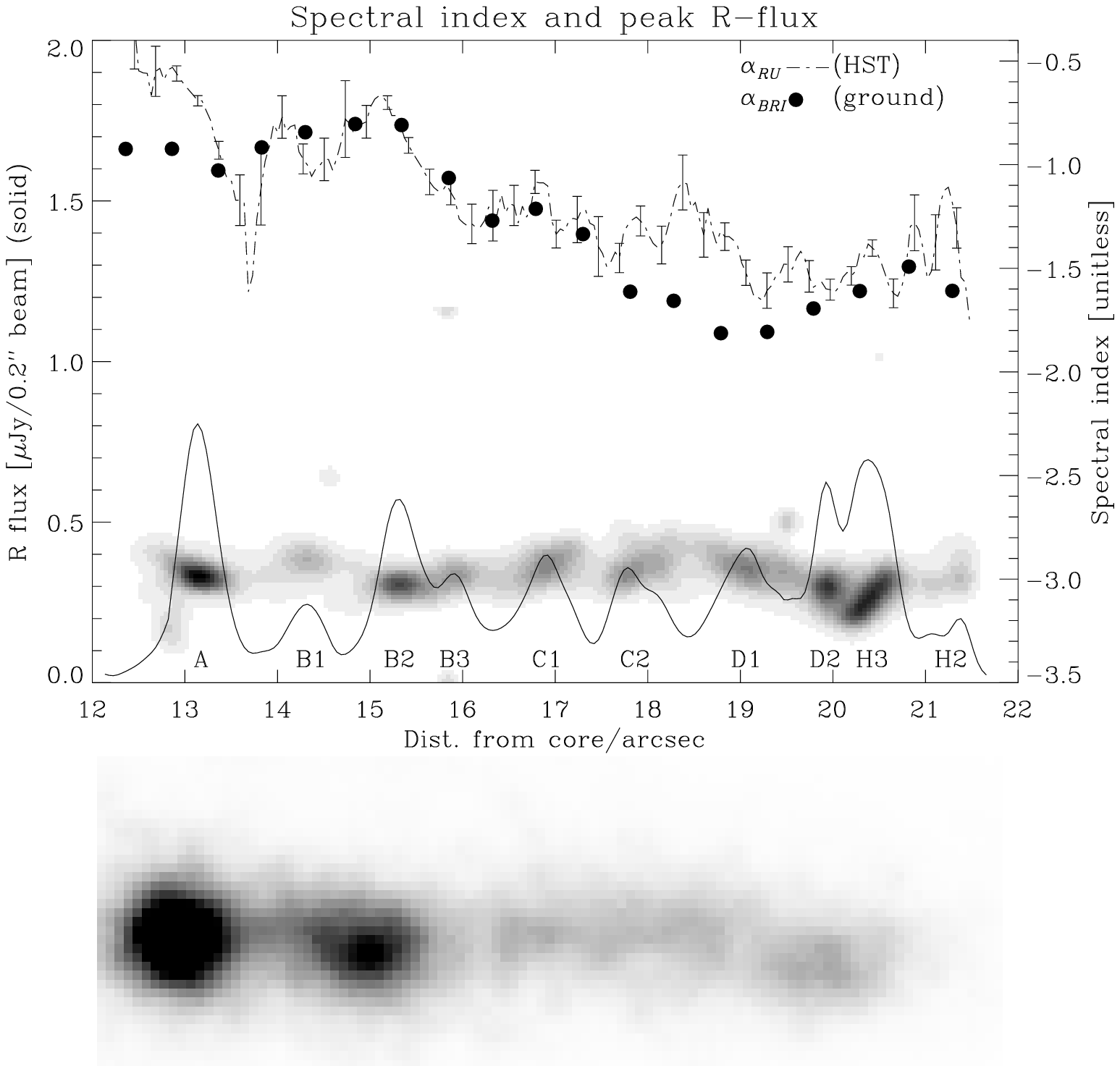,height=10cm}}}

\smallskip



{\noindent Figure 2. Optical (Jester et al. 2001b, top) and X-ray emissions
(bottom) of the 3C273 jet.  The top panel shows runs of $\alpha_o$ (points and
dashed curve at top) and optical flux (solid curve). The Chandra image is
adaptively smoothed. By contrast to M87, in the 3C273 jet optical flux and 
$\alpha_o$ are uncorrelated.  Also, the X-ray and optical maxima of knot B are
offset by $\sim 0.5''$ and X-ray emission is not seen at  $\theta_{nuc}\gsim
21''$.  This implies different physics than seen in the M87 jet.}

\section {Discussion}

There are significant differences between the observed morphologies of the M87
and 3C273 jet, which probably translate to differences in structure and
physics.  Unfortunately, not all of the observations are in place even for
these objects.  For 3C273, better HST polarimetry is required, and several more
bands of deep optical imaging would be very helpful to pin down the $\alpha_o$
map better (compare the error bars in the runs of $\alpha_o$ in Figures 1 and
2).  For M87, the magnetic and energetic structures are better constrained, but
there are other issues, for example the observed superluminal motion in several
knots (Biretta et al. 1999a,b)  and the evolution of physical conditions in
those components.

We are only scratching the surface of the range of properties in jets.  Many
new observations are needed.  Through cycle 9, four of the twelve known jets
brighter than $20$ mag/arcsec$^2$  have only snapshot HST observations, and
four more have only shallow exposures and poor spectral coverage.  Moreover,
through Cycle 9, polarimetry had been done only for M87 and 3C273 (3C264 and
3C78 are scheduled for Cycle 10).  Chandra is doing somewhat better; deep
observations have been obtained or scheduled for 7 of the 12 brightest optical
jets.

\end{document}